\documentclass{aa}
\usepackage{epsf}
\catcode`\"=\active\let"=\"
\def\3{\ss }

\newcommand{\hbeta}{H$\beta$}

\newcommand{\teff}{T_{\mathrm{eff}}}

\newcommand{\aj}{AJ}
\newcommand{\apj}{ApJ}
\newcommand{\apjs}{ApJS}

\newcommand{\aua}{A\&A}
\newcommand{\auas}{A\&AS}
\newcommand{\mnras}{MNRAS}

\newcommand{\pasp}{PASP}

\newcommand{\tkin}{t_{\mathrm{kin}}}

\begin{document}

\thesaurus{06(08.04.1; 08.06.3; 08.23.1; 09.16.1)}

\title{Spectroscopic investigation of old planetaries}
\subtitle{V.~Distance scales}
\author{R.~Napiwotzki
\thanks{Visiting astronomer, German-Spanish Astronomical Center, Calar Alto,
Spain, operated by the Max-Planck-Institut f\"ur Astronomie,
Heidelberg, jointly with the Spanish National Commission for Astronomy}
}
\institute{
Dr.~Remeis-Sternwarte, Sternwartstr.~7, D-96049~Bamberg, Germany
}
\offprints{Ralf Napiwotzki}
\date{Received date; accepted date}

\maketitle
\markboth{R.~Napiwotzki: Spectroscopic investigation
	of old planetaries  V.}
	{R.~Napiwotzki: Spectroscopic investigation
	of old planetaries  V.}

\begin{abstract}
We use the results of our recent NLTE model atmosphere analysis of
central stars of old planetary nebulae (PN) to calculate distances.
We perform a comparison with three other methods (trigonometric parallaxes,
interstellar Na\,D lines, and Shklovsky distances) and discuss the
problem of the PNe distance scale.
The result of the comparison of our spectroscopic distances 
with the trigonometric distances is that the spectroscopic distances are
55\% larger. Since using trigonometric parallaxes
with large relative measurement errors
can introduce systematic errors, we carried out a Monte Carlo simulation 
of the biases
introduced by selection effects and measurement errors.
It turns out 
that a difference between both distance scales of
the observed
size is expected for the present day data if the underlying distance
scales are identical. Thus our finding is essentially a confirmation of the
spectroscopic distance scale! Good agreement is found between the 
spectroscopic distances and distances derived from the interstellar
NaD lines. 

All three independent methods of distance measurement
indicate that the widely used ``statistical'' distance scales of the
Shklovsky type are too short for old PNe. A correlation with nebular radii
exists. The most likely explanation
is an underestimate of the nebula masses for large PN. Implications for the
nebula masses are discussed. Estimates of the PNe space
density and birthrate, which are based on Shklovsky type distances,
therefore give too large values. 

\keywords{stars: distances --- planetary
	nebulae: general --- white dwarfs --- stars: fundamental parameters}

\end{abstract}

\section{Introduction}
Distances of PNe are usually difficult to determine and are a long
standing problem of PNe investigations.
The vast majority of published distances of PNe are based on the
Shklovsky (\cite{Shk56}) method or derivatives of it (often called statistical
distances). This method allows the
calculation of PN distances from the measurement of the recombination
line \hbeta\ and the angular diameter.  
However, Shklovsky distances are notoriously smaller than distances 
derived from
model atmosphere analysis of the central stars (cf.\ M\'endez \cite{MKH88}). 

The question of PNe space densities and birth rates is closely related
to the distance scale problem.
Ishida \& Weinberger (\cite{IW87}) 
compiled a list of nearby PNe, which contains mostly old,
evolved nebulae (actually many of our PNe were selected from this list). 
Ishida \& Weinberger collected distance determinations from literature and 
computed the space density and birth rates of this local sample of PNe. 
The derived birth rate of $8\cdot 10^{-12}\,\mathrm{yr^{-1}pc^{-3}}$ is 
too high to be in accordance with estimates of 
white dwarf birth rates ($2.3\cdot
10^{-12}\,\mathrm{yr^{-1}pc^{-3}}$; Weidemann \cite{Wei91}). 
Since every central star of a PN (CSPN) should become a 
white dwarf this yielded a real dilemma. Taken at face value this
would indicate that current white dwarf samples are very incomplete
and the white dwarf birth rates are seriously underestimated. A
certain fraction of white dwarfs may be hidden in binaries, indeed. 
Weidemann (\cite{Wei91}) used a very local sample of white dwarfs ($d<10$\,pc)
for his estimate and applied corrections for incompleteness and binarity.
Pottasch (\cite{Pot96}) reevaluated the PN space density and
derived a value of $3\cdot 10^{-12}\,\mathrm{yr^{-1}pc^{-3}}$, lower 
than Ishida \& Weinberger's result, but still higher than the
estimated white dwarf birthrate. The difference in the PNe birthrate
can be traced back to the use of different distance scales. 
While Ishida \& Weinberger's collection is mainly based on statistical distance
estimates (mainly Shklovsky distances and derivates of this method),
Pottasch excluded statistical distance determinations.

In Paper~IV of this series (Napiwotzki \cite{Nap99}) 
we presented the results of an NLTE
model atmosphere analysis of 27 central stars of old PNe.
This analysis of a reasonably sized sample of central stars of old PNe 
enables us
to address the question of the distance scale of these objects. 
In Paper~III (Napiwotzki \& Sch\"onberner \cite{NS95})
we proposed the use of the interstellar Na\,D lines for distance
determinations. These distances are free of assumptions about the
nebula or the central star and are on average larger than the
Shklovsky distances by a factor of 2.5 for our sample of old PNe. 
During the last years a
number of trigonometric parallax measurements became available for a
sample of central stars of old PNe (Harris et al.\ \cite{HDM97}, 
Pottasch \& Acker \cite{PA98}, Guti\'errez-Moreno et al.\ \cite{GAL99}). 
We will show that the CSPNe distances
derived form our model atmosphere analysis, the Na\,D distances, and
the trigonometric parallaxes are in agreement, but all three
distances scales are much larger than those based on the Shklovsky method. 

\section{Distance scales}

\subsection{Spectroscopic distances}

\begin{table}
\caption{Model atmosphere fluxes 
(in $10^8$erg\,cm$^{-2}$s$^{-1}$\AA$^{-1}$) at 
$\lambda_{\mathrm{eff}} = 5454$\,\AA\
for the calculation of absolute magnitudes and distances. These values
are calculated for a hydrogen-rich composition
($n_{\mathrm{He}}/n_{\mathrm{H}} = 0.1$) and $\log g = 6.5$. However,
the $V$ band flux is quite insensitive to composition and $\log g$}
\label{t:fmod}
\begin{tabular}{rr|rr}
$T_{\mathrm{eff}}$/K  &$F_{5454}$
&$T_{\mathrm{eff}}$/K  &$F_{5454}$\\ \hline
40000     &2.18   &120000    &6.37\\
50000     &2.76   &140000    &7.21\\
60000     &3.31   &170000    &8.50\\
70000     &3.87   &200000    &9.67\\
80000     &4.51   &250000    &11.93\\
90000     &5.05   & 300000    &13.98\\
100000    &5.52   &&\\
\end{tabular}
\end{table}

Spectroscopic distances can be derived from the model atmosphere
analyses described in Paper~IV and offer an approach to the PN
distance scale independent from the properties of the nebulae.
After the stellar mass is estimated from a comparison with
evolutionary models (cf.\ Paper~IV) the distance can be calculated
in a straightforward way from effective temperature, gravity, and the 
dereddened apparent magnitude of
the stars:
\begin{enumerate}
\item the stellar radius is computed from surface gravity and mass,
\item the surface flux in the $V$ band is calculated from model atmospheres,
   with the appropriate parameters (thus bypassing the uncertain bolometric
   corrections),
\item and finally the absolute magnitude $M_V$ is derived and compared with the
   dereddend $V_0$ magnitude of the star.
\item This procedure can be condensed into one formula:
   \begin{equation}
   d = 0.111 \sqrt{\frac{M_\star F_V}{g}10^{0.4V_0}}
   \end{equation}
   where $M_\star$ is the solar mass in $M_\odot$, $g$ is the surface gravity
   in $\mathrm{cm}/\mathrm{s}^2$ and $F_V$ is the model atmosphere
   flux in $10^8$erg\,cm$^{-2}$s$^{-1}$\AA$^{-1}$. We use the monochromatic
   approximation of Heber et al.\ (\cite{HHJ84}) for $F_V$ with an effective
   wavelength $\lambda_{\mathrm{eff}} = 5454$\,\AA, appropriate for hot stars.
   A short tabulation of model atmosphere fluxes is provided in
   Table~\ref{t:fmod}.
\end{enumerate}
This method is independent of any assumption about the PN,
but obviously relies on the quality of the model atmospheres used for
the analysis and, to a lesser extent, on the evolutionary models used
for the mass determination. For our objects 
the apparent magnitudes were taken from Table~4 of
Paper~III. If a high quality measurement (indicator ``A'') exists in
the literature, it
was adopted. Otherwise our spectrophotometric measurements were
used. Since only very few reliable extinction measurements are
available for our objects, we estimated the reddening from the
galactic interstellar extinction model of Arenou et al.\
(\cite{AGG92}). However, reddening is generally small and therefore only
of moderate importance for our conclusions. The resulting distances
are listed as $d_{\mathrm{NLTE}}$ in Table~\ref{t:dist} together with
the analysis results of Paper~IV. Error limits
were estimated from the analyses errors given in Paper~IV, 
the estimated errors of the $V$ measurements, and the reddening. The
dominant error source is in most cases the gravity determination.

\begin{table*}
\caption{Comparison of different distance determinations: from 
our NLTE analysis in Paper~IV $d_{\mathrm{NLTE}}$, 
trigonometric distances $d_\Pi$ of Harris et al.\ (\cite{HDM97}),
from the interstellar Na\,D line $d_{\mathrm{Na\,D}}$, and the
Shklovsky distances $d_{\mathrm{Shkl}}$ (the latter two are taken 
from Table~6 of Paper~III). The kinematical ages calculated ($\tkin$) from 
the spectroscopic distances are also provided. Effective temperature
$\teff$, surface gravity $g$, and stellar mass $M$ were taken from
Table~1 of Paper IV. For further input data cf.\
Table~6 of Paper~III}

\label{t:dist}
\begin{tabular}{ll|rrr|r@{}llr@{}l|r@{}lrr}
PN\,G	&name	
	&$\teff$	&$\log g$	&$M$ 
	&\multicolumn{2}{c}{$d_{\mathrm{NLTE}}$}
	&$R$	&\multicolumn{2}{c}{$\tkin$}
	&\multicolumn{2}{c}{$d_\Pi$}	
	&$d_{\mathrm{Na\,D}}$	&$d_{\mathrm{Shkl}}$ \\
	&	&(K)	&($\mathrm{cm\,s^{-2}}$)	&($M_\odot$)
	&\multicolumn{2}{c}{(pc)} 
	&(pc)	&\multicolumn{2}{c}{($10^3$yrs)}
	&\multicolumn{2}{c}{(pc)}	
	&(pc)	&(pc) \\ \hline
025.4$-$04.7	&\object{IC\,1295}	&90100	&6.66	&0.51
	&700&$^{+400}_{-300}$	&0.16	&7&.6	
	&&$$			&1500	&1000	\\
027.6+16.9	&\object{DeHt\,2}	&117000	&5.64	&0.64
	&2400&$^{+760}_{-600}$	&0.54	&27&
	&&$$			&	&1900	\\
030.6+06.2	&\object{Sh\,2-68}	&95800	&6.78	&0.55
	&1100&$^{+500}_{-400}$	&1.02	&200&	
	&&$$			&600	&310	\\
034.1$-$10.5	&\object{HDW\,11}	&68100	&6.38	&0.39
	&1200&$^{+500}_{-400}$	&0.13	&6&.6	
	&&$$			&1000	&4500	\\
036.0+17.6	&\object{A\,43}		&116900	&5.51	&0.68
	&2600&$^{+800}_{-700}$	&0.51	&25&	
	&&$$			&	&1600	\\
036.1$-$57.1	&\object{NGC\,7293}	&103600	&7.00	&0.57
	&290&$^{+90}_{-70}$	&0.69	&28&
	&210&$^{+40}_{-30}$	&	&160	\\
047.0+42.4	&\object{A\,39}		&117000	&6.28	&0.57
	&1900&$^{+600}_{-500}$	&0.81	&22&
	&&$$			&	&1200	\\
060.8$-$03.6	&\object{NGC\,6853}	&108600	&6.72	&0.56
	&440&$^{+150}_{-120}$	&0.42	&13&
	&380&$^{+80}_{-50}$	&550	&260	\\
063.1+13.9	&\object{NGC\,6720}	&101200	&6.88	&0.56
	&1100&$^{+400}_{-300}$	&0.20	&2&.6	
	&700&$^{+450}_{-200}$	&990	&870	\\
066.7$-$28.2	&\object{NGC\,7094}	&125900	&5.45	&0.87
	&2200&$^{+700}_{-600}$	&0.51	&11&
	&&$$			&	&1400	\\
072.7$-$17.1	&\object{A\,74}		&108000	&6.82	&0.56
	&1700&$^{+700}_{-500}$	&4.52	&170&
	&750&$^{+650}_{-240}$	&750	&260	\\
077.6+14.7	&\object{A\,61}		&88200	&7.10	&0.55
	&1400&$^{+800}_{-500}$	&0.67	&22&
	&&$$			&	&850	\\
111.0+11.6	&\object{DeHt\,5}	&76500	&6.65	&0.44
	&510&$^{+170}_{-140}$	&0.66	&130&
	&&$$			&440	&400	\\
120.3+18.3	&\object{Sh\,2-174}	&69100	&6.70	&0.43
	&560&$^{+140}_{-110}$	&0.81	&40&
	&&$$			&500	&	\\
124.0+10.7	&\object{EGB\,1}		&147000	&7.34	&0.65
	&650&$^{+290}_{-210}$	&0.41	&20&
	&&$$			&480	&610	\\
125.9$-$47.0	&\object{PHL\,932}	&35000	&5.93	&0.28
	&240&$^{+50}_{-40}$	&0.16	&7&.7
	&110&$^{+50}_{-30}$	&	&820	\\
128.0$-$04.1	&\object{Sh\,2-188}	&102000	&6.82	&0.56
	&1000&$^{+1000}_{-600}$	&0.88	&22&	
	&&$$			&600	&220	\\
148.4+57.0	&\object{NGC\,3587}	&93900	&6.94	&0.55
	&1300&$^{+600}_{-400}$	&0.52	&13&
	&&$$			&	&620	\\
149.4$-$09.2	&\object{HDW\,3}		&125000	&6.75	&0.58
	&1500&$^{+800}_{-600}$	&1.90	&93&
	&&$$			&1100	&410	\\
156.3+12.5	&\object{HDW\,4}		&47300	&7.93	&0.64
	&250&$^{+80}_{-70}$	&0.063	&3&.1
	&&$$			&	&2000	\\
156.9$-$13.3	&\object{HaWe\,5}	&38100	&7.58	&0.51
	&420&$^{+140}_{-120}$	&0.035	&1&.7
	&&$$			&	&6800	\\
158.5+00.7	&\object{Sh\,2-216}	&83200	&6.74	&0.49
	&190&$^{+50}_{-40}$	&2.69	&660&
	&130&$^{+9}_{-8}$	&110	&30	\\
158.9+17.8	&\object{PuWe\,1}	&93900	&7.09	&0.56
	&700&$^{+250}_{-200}$	&2.02	&73&
	&430&$^{+90}_{-60}$	&470	&140	\\
197.4$-$06.4	&\object{WeDe\,1}	&141000	&7.53	&0.68
	&1000&$^{+500}_{-300}$	&2.17	&130&
	&&$$			&880	&320	\\
204.1+04.7	&\object{K\,2-2}		&67000	&6.09	&0.38
	&630&$^{+260}_{-210}$	&0.66	&65&
	&&$$			&1200	&440	\\
215.5$-$30.8	&\object{A\,7}		&99000	&7.03	&0.57
	&700&$^{+500}_{-300}$	&1.30	&63&
	&$\ge 700$&		&	&220	\\
219.1+31.2	&\object{A\,31}		&84700	&6.63	&0.48
	&1000&$^{+500}_{-400}$	&2.32	&65&
	&210&$^{+110}_{-50}$	&	&230	\\
\end{tabular}
\end{table*}

\subsection{Trigonometric distances}

\begin{figure}
\epsfxsize=8.5cm
\epsffile[18 170 592 690]{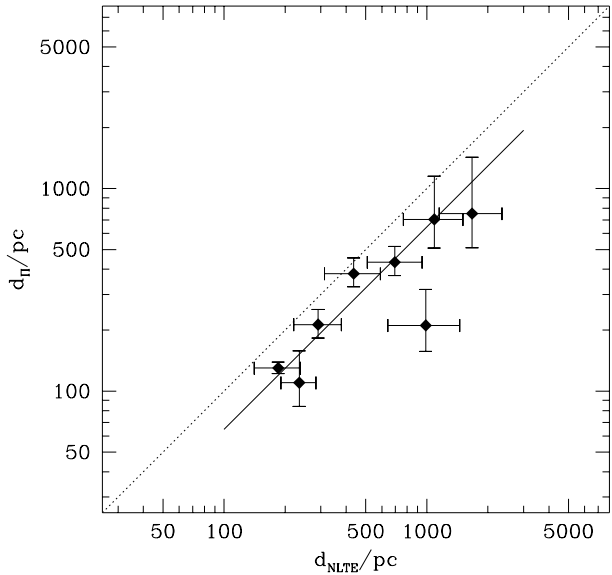}
\caption[]{Distances computed from the results of the NLTE analysis
compared to trigonometric distances. 
The error bars correspond to the measurement
errors given in Table~\ref{t:dist}. 
The solid line indicates the average ratio of the
NLTE distances and the trigonometric distances, the 
dotted line equality}
\label{f:dtrig}
\end{figure}

Recently Harris et al.\ (\cite{HDM97}) published trigonometric
parallaxes of 11 CSPNe. Eight of these stars belong to our sample (the
value given for \object{A\,7} is actually only an upper limit). 
Pottasch \& Acker (\cite{PA98}) 
discussed HIPPARCOS parallaxes of three CSPNe including
\object{PHL\,932} which belongs to our sample.
The resulting
distance values are listed as $d_\Pi$ in Table~\ref{t:dist} and
are compared with the model atmosphere results in Fig.~\ref{f:dtrig}. The
error limits were computed from the measurement errors of the
parallaxes. 

As it is evident from Fig.~\ref{f:dtrig} and the following figures 
the error distribution of
distance determinations is highly non-Gaussian with many outliers,
which can skew the common weighting techniques to erroneous
values. The reason is that the usually (implicitly) adopted Gaussian
probability distribution gives high weight to deviant points.
Therefore we decided 
to minimize the
absolute deviations, which corresponds to a double sided exponential
probability distribution and provides a more robust estimate 
(see discussion in Press et al.\ \cite{PTV92}). 
Since the distance errors are highly
asymmetric, we did the comparison with the parallaxes, which have
roughly symmetric error limits. Since we in all cases compare two measurements
suffering from large uncertainties, 
we performed pro forma a linear regression with
allowance for errors in both directions and with the intersection fixed
at zero.
The error ranges given below correspond to the $1\sigma$
error of the mean.

The measured trigonometric distances are always
smaller than the
NLTE distances (cf.\ Fig.~\ref{f:dtrig}).
The weighted mean of the distance
ratios $r_{\mathrm{obs}}$, computed as described above,  amounts to 
\begin{equation}
r_{\mathrm{obs}} = \frac{d_{\mathrm{NLTE}}}{d_\Pi} =  1.55\pm 0.29 
\label{eq:Shkl}
\end{equation}
Taken at face value this in concordance with the conclusions of 
Jacoby \& Van de Steene (\cite{JVS95}), who compared trigonometric parallax
measurements of Pier et al.\ (\cite{PHD93}) with spectroscopic distances and
concluded that the model atmosphere results overestimate the distances
by $\approx$40\%. 
However, one must consider that trigonometric parallaxes are on the
one hand the most direct method for distance determinations, but on
the other hand are also subject to heavy biases,
which are introduced by random errors of measurement and on average
cause the trigonometric parallaxes to be overestimated.
This was discussed as early as \cite{TW53} by Trumpler \& Weaver. Lutz \&
Kelker (\cite{LK73}) were the first to provide a quantitative correction,
which is a function of the relative measurement error
of the parallax $\sigma_\Pi /\Pi$. 
The original Lutz-Kelker correction
is valid in a strict sense only,  if nothing else is known about the
star, which is usually an unrealistic assumption. Different kinds if
corrections have to be applied if one deals with a volume limited sample,
a magnitude limited sample, or if the ``luminosity function'' of the parent
population is known (Lutz \cite{Lut83}; Smith \cite{Smi87}).

We took a different approach and performed a Monte Carlo simulation to derive
biases caused by the selection of CSPNe for parallax measurements and the
accompanying measurement errors. Our working hypothesis is that the
spectroscopic distance scale is essentially correct, i.e.\ no systematic
errors are present. The results of the Monte Carlo simulations 
are used to test if this hypothesis is compatible with observations.

\begin{figure}
\epsfxsize=8.5cm
\epsffile[18 40 750 580]{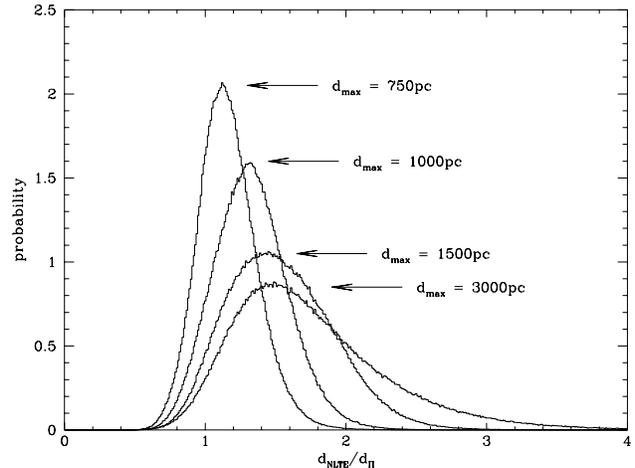}
\caption[]{Results of Monte Carlo simulations for several values of the
maximum allowed ``spectroscopic'' distance $d_{\mathrm{max}}$. The 
distributions are normalized to an integrated area of 1. For cosmetic 
reasons we used
$N_{\mathrm{total}}=10^6$ for this plot}
\label{f:mc_dmax}
\end{figure}

In our simulation CSPNe were randomly created according to a simple model of
the Galactic stellar density distribution and a theoretical post-AGB track.
A spectroscopic analysis is simulated, i.e.\ random measurement errors are
added. The star is selected for parallax measurement if the ``spectroscopic
distance'' is below a threshold value $d_{\mathrm{max}}$. A measurement error
of the parallax $\sigma_\Pi$ is added to the true parallax and the resulting
distance used for a comparison with the spectroscopic parallax, if the
``measured parallax'' is larger than a threshold value $\Pi_{\mathrm{min}}$.
$N_{\mathrm{sample}}$ stars are collected and the mean ratio
$r_{\mathrm{MC}} = \frac{d_{\mathrm{NLTE}}}{d_\Pi}$ 
is computed analogous to the procedure
applied for the observed data. This process is repeated 
$N_{\mathrm{total}}$ times
and a probability distribution is evaluated. Sample probability
distributions are given in Fig.~\ref{f:mc_dmax} (these are discussed below).
A detailed description of the Monte Carlo simulations, the input parameters,
and their standard values is given in the Appendix. 

An investigation of the influence of the different input parameters on the
results can be found in Table~\ref{t:mcvar}. The derived bias between
spectroscopic and trigonometric distances is most sensitive to the chosen
threshold value $d_{\mathrm{max}}$ for the pre-selection of stars, the
parallax measurement error $\sigma_\Pi$ and the threshold parallax
$\Pi_{\mathrm{min}}$. Our standard model assumes $d_{\mathrm{max}} = 1000$\,pc.
Note that one star in the Harris et al.\ (\cite{HDM97}) 
sample has a much larger
value of the spectroscopic distance (\object{A\,74}; 
$d_{\mathrm{NLTE}} = 1700$\,pc).
For $\sigma_\Pi$ the mean of the error limits in Table~\ref{t:dist},
$\sigma_\Pi = 0.95$, was adopted. The threshold parallax was set to
$\Pi_{\mathrm{min}} = 1.33$ (corresponds to \object{A\,74}). 
Our standard sample size,
$N_{\mathrm{sample}} = 8$, corresponds to the observed sample and the 
simulation is carried out for $N_{\mathrm{total}} =100000$ samples.

The first qualitative conclusion, which can be drawn from
Table~\ref{t:mcvar} and Fig.~\ref{f:mc_dmax} is that the ``measured'' ratio
$r_{\mathrm{MC}}$ is {\em always considerably larger than 
unity}, although the underlying distance scales used in the Monte Carlo
simulations are identical. Our standard model yields
\begin{equation}
r_{\mathrm{MC}} = 1.32 \pm 0.25
\end{equation}
(we give the median value and the range which contains 68.3\% of all samples
as $1\sigma$ limit). Although our measured value
$r_{\mathrm{obs}} = 1.55\pm 0.29$ is larger it is still in the
$1\sigma$ range of the standard simulation.
If we increase $d_{\max}$ to 1500\,pc (keeping \object{A\,74} in mind) 
the result of
the Monte Carlo simulation increases to
$ r_{\mathrm{MC}} = 1.50 \pm 0.38$
close to the measured value.
The impact of modifying $d_{\max}$ on the probability distribution
is displayed in Fig.~\ref{f:mc_dmax}.
If we on the other hand remove CSPNe with $d_{\mathrm{NLTE}}>1000$\,pc from our
sample in Table~\ref{t:dist} (\object{NGC\,6720} and \object{A\,74}) we derive
$ r_{\mathrm{MC}} = 1.42 \pm 0.29$
close to the value predicted for $d_{\max}=1000$\,pc.

A paradoxical effect results if the threshold parallax $\Pi_{\mathrm{min}}$ is
increased: the bias increases for increasing $\Pi_{\mathrm{min}}$, 
if the spectroscopic threshold distance is left unchanged! E.g.\ for 
$\Pi_{\mathrm{min}}=3.0$\,mas we derive
$r_{\mathrm{MC}} = 1.51\pm 0.35$. The reason is that we
introduce a strong selection effect for stars with much too large measured
parallaxes. This effect is only overcome, if $\Pi_{\mathrm{min}}$ is increased
to even larger values ($\Pi_{\mathrm{min}}\ge5$\,mas).

In principle, the expected bias can be reduced by lowering $d_{\mathrm{max}}$
and $\sigma_\Pi$. However, if we e.g.\ adopt $d_{\mathrm{max}} = 750$\,pc
and $\sigma_\Pi = 0.7$\,mas our observed sample is reduced to 3 stars!

We conclude that the mean value of $r_{\mathrm{MC}}$ derived
from our observed sample (Table~\ref{t:dist}) is well within the range
predicted by Monte Carlo simulations for perfect agreement of both distance
scales. Thus the trigonometric parallax measurements provide no evidence
that the spectroscopic distances are in error, but confirm the spectroscopic
distance scale, instead! 
However, due to the statistical uncertainties we cannot
provide a definitive proof of an agreement of better than about 20\%.

\subsection{Distances based on interstellar lines and extinction}

Another method independent of assumptions about central star or nebula
properties is based on interstellar lines. In Paper~III we used 
the interstellar Na\,D lines at 5890/96\,\AA\ for this purpose. The
CSPNe are too hot to show any photospheric Na\,D lines and the PNe
material is extremely dispersed. In their survey Dinerstein et al.\
(\cite{DSU95}) detected circumstellar Na\,D lines in nine PNe. However, all
investigated PNe were young objects, and if there is any neutral sodium at all
in the circumstellar matter of the central star of old PNe, the column density
would be very low and the resulting nebula contribution negligible.
Naturally this method is restricted to stars at low galactic
latitude. The distances derived in paper~III are given as
$d_{\mathrm{Na\,D}}$ in Table~\ref{t:dist}. 

Our Na\,D distances in paper~III were determined from the map of the
interstellar Na\,D line strength of Binnendijk (\cite{Bin52}). 
One might ask, whether
the distance scale adopted in this work is still valid. To our knowledge
no more recent collection of interstellar Na\,D equivalent widths is available.
The reason is that using the equivalent width is out of fashion, because 
resolved interstellar lines in high resolution spectra provide more
information. However, for our faint CSPNe equivalent widths are still a useful
tool. 

Binnendijk (\cite{Bin52}) adopted spectroscopic distances of B stars, 
which were
determined by the Yerkes group (Ramsey \cite{Ram50}, 
Duke \cite{Duk51}, and unpublished
distances provided by W.W.~Morgan). It is save to assume that systematic
differences between these authors are small. We performed a check of the
Ramsey (\cite{Ram50}) and Duke (\cite{Duk51}) distances. 
For this purpose we selected a
representative subsample of 20 stars from each collection 
with Str\"omgren $uvby\beta$ measurements in the Hauck \& Mermilliod
(\cite{HM98}) catalogue.
Temperatures and gravities were calculated with an updated version
(Napiwotzki \& Lemke, in prep.) of the photometric calibration of Napiwotzki
et al.\ (\cite{NSW93}). The new gravity calibration is based on accurate
trigonometric parallaxes measurements of B stars with HIPPARCOS. Since the
calibration doesn't cover O stars and supergiants, these stars were excluded
from the comparison. Masses were derived by interpolation in the
evolutionary tracks of Schaller et al.\ (\cite{SSM92}) and distances 
$d_{uvby\beta}$
computed as described in Section~2.1. Finally these distances were compared
with the spectroscopic distances provided by Ramsey (\cite{Ram50}) and Duke
(\cite{Duk51})
and mean ratios were computed as described in Sect.~2.2. The results are
\begin{eqnarray}
\frac{d_{uvby\beta}}{d_{\mathrm{Ramsey}}} &= &0.92 \\
\frac{d_{uvby\beta}}{d_{\mathrm{Duke}}} &= &1.17
\end{eqnarray}
or in other words the Ramsey distances are on average 8\% longer, the Duke
distances 14\% shorter than the $uvby\beta$ distances. 

Although the scatter of the individual determinations is large we can
conclude that altogether systematic differences  between the old Yerkes 
system and modern photometric distances, which are tied to accurate HIPPARCOS 
parallaxes, are small.

\begin{figure}
\epsfxsize=8.5cm
\epsffile[18 170 592 690]{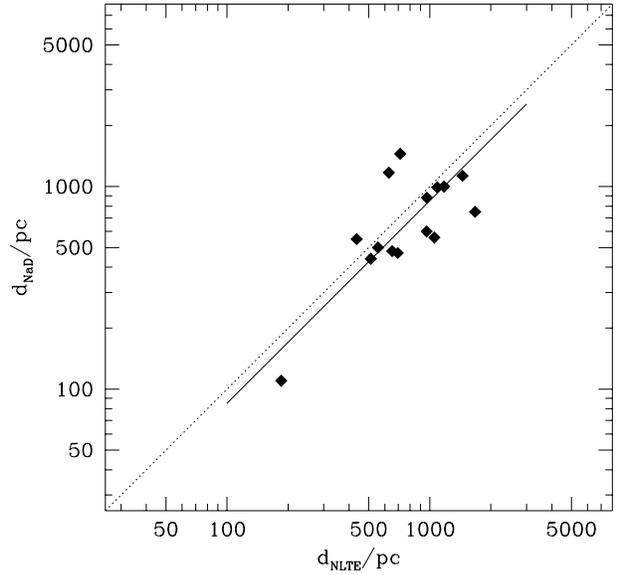}
\caption[]{Distances computed from the results of the NLTE analysis
compared to Na\,D distances. The solid line indicates the average
distance ratio}
\label{f:dnad}
\end{figure}

Fig.~\ref{f:dnad} shows
that the agreement of the Na\,D distances with the model atmosphere
analysis is good. The scatter is easily explained by the spatially highly
variable extinction in the galactic plane. The average ratio of both
distance scales amounts to 
\begin{equation}
\frac{d_{\mathrm{NLTE}}}{d_{\mathrm{Na\,D}}} = 1.18\pm 0.22\ .
\end{equation}

The extinction distance method is an approach similar to our Na\,D
method. The interstellar reddening of the central star or the PN is
measured and compared to a distance-reddening relation derived from
stars in the angular vicinity of the PN. Since reddening and distances
of many stars have to be determined, it is unfortunately a costly
method and has
been applied to few PNe only. Extinction distances of two PNe of our
sample were measured by Saurer et al.\ (\cite{Sau95}). Results are $800\pm
300$\,pc for \object{Sh\,2-188} and $800\pm 400$\,pc for \object{HDW\,3}. 
The \object{Sh\,2-188}
distance is in good agreement with our estimate of
$1000^{+1000}_{-600}$\,pc. Our HDW\,3 distance of
$1500^{+800}_{-600}$\,pc is moderately larger than the extinction
distance. Saurer noted that his upper limit maybe uncertain because the
interstellar extinction reaches a plateau value at 1200\,pc. However,
the error bars overlap anyway. 

\subsection{Shklovsky distances}

The Shklovsky (\cite{Shk56}) method allows the
calculation of PN distances from the measurement of the recombination
line \hbeta\ and the angular diameter. The distance can be computed from
(see e.g.\ Pottasch \cite{Pot84}, p.~115): 
\begin{equation}
d_{\mathrm{Shkl}} = 218 \frac{M^{0.4}_{\mathrm{ion}} t^{-0.18}}
	{\epsilon^{0.2} F^{0.2}_{\mathrm{H}\beta}\Theta^{0.6}}
\end{equation}
with distance $d_{\mathrm{Shkl}}$ in pc, flux $F_{\mathrm{H}\beta}$
in erg\,cm$^{-2}$s$^{-1}$, the angular diameter 
$\Theta$ in arc sec, the electron
temperature $t$ in $10^4$\,K, the filling factor
$\epsilon$, and the mass of the ionized nebular matter
$M_{\mathrm{ion}}$ in $M_\odot$. The dependence of $d_{\mathrm{Shkl}}$
on $t$ and $\epsilon$ is only small, hence it suffices to use standard
values (e.g.\ $t = 1$ and $\epsilon = 0.75$). 
However, major problems are caused by the need to 
assume a mass $M_{\mathrm{ion}}$ of the (ionized) nebula. The
classical Shklovsky method adopts a constant mass, usually
$M_{\mathrm{ion}} = 0.2 M_\odot$. However, for the few PNe with
independent  distance determinations a trend of increasing mass with
increasing radius is observed. (see e.g.\ Maciel \& Pottasch \cite{MP80};
Cahn et al.\ \cite{CKS92}). 
Thus Maciel \& Pottasch (\cite{MP80}), Daub (\cite{Dau82}), and
Cahn et al.\ (\cite{CKS92}) proposed modified Shklovsky methods employing an
empirical mass-radius relationship. However, the Daub (\cite{Dau82}) and Cahn
et al.\ (\cite{CKS92}) mass-radius relations
adopted an upper limit for $M_{\mathrm{ion}}$, which means that for
old, large PNe a constant mass of the order $0.2 M_\odot$ is applied,
similar to the 
classical version. Recent approaches by Zhang (\cite{Zha95}) and by Van de
Steene \& Zijlstra (\cite{VSZ95}) propose to use the correlation between the 
radio continuum brightness temperature and radius instead. However,
this method needs an empirical calibration, too. In Table~\ref{t:dist}
we present Shklovsky distances $d_{\mathrm{Shkl}}$ collected from the
literature. Most values were taken from the catalogue of Cahn et al.\
(\cite{CKS92}). Please, consult Table~6 of paper~III for detailed references. 

\begin{figure}
\epsfxsize=8.5cm
\epsffile[18 170 592 690]{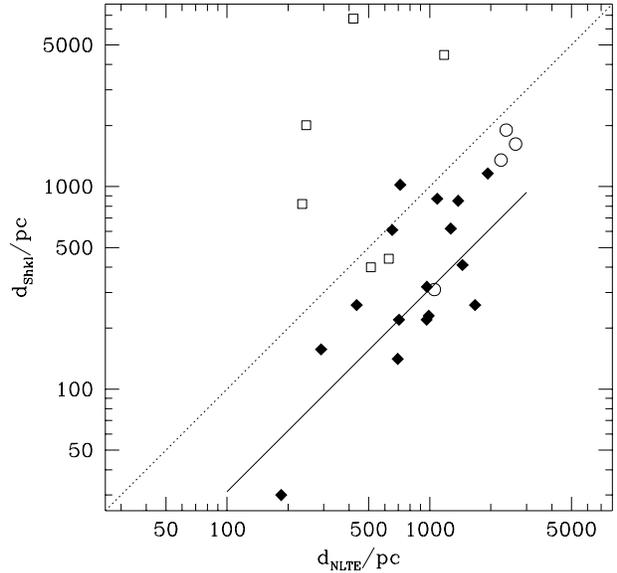}
\caption[]{Distances computed from the results of the NLTE analysis
compared to Shklovsky distances. Filled diamonds: ordinary white
dwarf CSPNe of spectral type DA and DAO, 
open circles: hybrid/high luminosity objects, open squares:
non post-AGB objects. The
dotted line indicates equality, the solid the average distance ratio
for the ordinary white dwarf CSPNe}
\label{f:dshkl}
\end{figure}

A comparison of the Shklovsky and the model atmosphere distances is
shown in Fig~\ref{f:dshkl}. We distinguish between ``ordinary''
white dwarf CSPNe of spectral type DA and DAO (filled symbols) 
and the hybrid/high luminosity objects
\object{A\,43}, \object{NGC\,7094}, \object{Sh\,2-68}, and \object{DeHt\,2}
(open circles) and the non post-AGB
objects (open squares). The evolutionary history of both latter
classes are likely very different from standard evolution and may
cause very different PN properties. The comparison shows that almost
all Shklovsky distances of the ordinary white dwarf CSPNe are smaller
than the model atmosphere distances. The average (weighted by the
error limits of our analysis) amounts to 
\begin{equation}
\frac{d_{\mathrm{NLTE}}}{d_{\mathrm{Shkl}}} = 3.2 \pm 1.3
\end{equation}
where we have adopted an $1\sigma$ error of all Shklovsky distances of
30\%.

Since $d_{\mathrm{NLTE}}/d_{\mathrm{Shkl}}$ is directly related to
the ionized mass of the PN by the Shklovsky formula $M_{\mathrm{ion}} \sim
d^{2.5}$, we interpret this as evidence that the ionized masses of the old PNe
are much higher than the adopted typical mass of $0.2 M_\odot$. The
distance scales would be in agreement if we increase the adopted PNe
mass to $3.5 M_\odot$. However, the large scatter is a first
indication that this is an
oversimplification of the real situation, as we will show below.
Note in passing that the non post-AGB objects (open squares in
Fig.~\ref{f:dshkl}) have considerably lower values of  
$d_{\mathrm{NLTE}}/d_{\mathrm{Shkl}}$ (mostly $<1$) indicating
masses lower than the adopted PN mass.  

\section{Discussion}

We have now shown that the distances $d_{\mathrm{NLTE}}$ derived from
our NLTE model atmosphere analyses are (within today's error limits)
in good agreement with the trigonometric parallaxes measured by 
Harris et al.\ (\cite{HDM97}) and Pottasch \& Acker (\cite{PA98}) 
after biases are taken into account,
and with distances derived from the strength of
the interstellar Na\,D lines. Both distance scales are model
independent and thus demonstrate that the NLTE analysis are not
subject to large systematic errors. On the other hand
Napiwotzki et al.\ (\cite{NGS99}) 
have shown that state-of-the-art analyses of hot 
white dwarfs performed independently by different groups can yield
surface gravities, which differ systematically by up to 0.1\,dex
which translates into a distance error of 12\%. Such errors of the
model atmosphere distance scale would  be compatible with the
trigonometric and Na\,D distances of our stars. Much larger
systematic errors can be excluded. The
Shklovsky distance scale of old PNe is shown to be too short,
most likely caused
by an underestimate of the PNe masses. 

\begin{figure}
\epsfxsize=8.5cm
\epsffile[18 40 750 580]{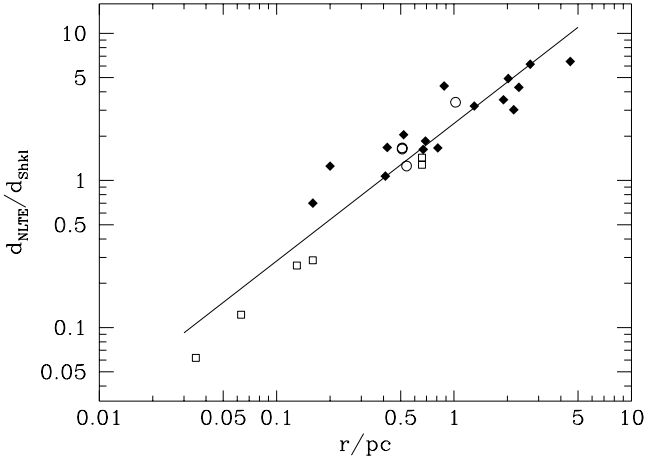}
\caption[]{The individual ratios of $d_{\mathrm{NLTE}}$ and
$d_{\mathrm{Shkl}}$ as function of the nebula radius. The meaning of
the symbols is the same as in Fig.~\ref{f:dshkl}. The line indicates
our fit (Eq.~\ref{eq:fit})}
\label{f:rcorrelation}
\end{figure}

\begin{figure}
\epsfxsize=8.5cm
\epsffile[18 40 750 580]{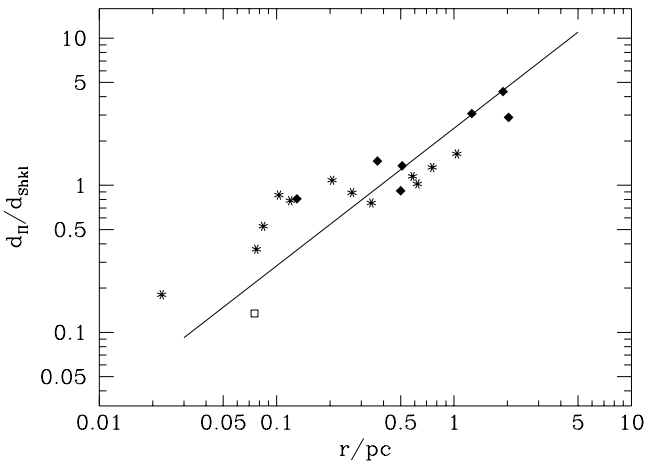}
\caption[]{Sames as Fig.~\ref{f:rcorrelation} but this time we compare 
the distances from trigonometric and spectroscopic parallaxes 
$d_{\mathrm{\Pi}}$ with $d_{\mathrm{Shkl}}$. The Ciardullo et al.\ (1999) 
results are plotted as asterisks. For other symbols cf.\
Fig.~\ref{f:dshkl}. The line indicates
the fit from Fig.~\ref{f:rcorrelation}}
\label{f:trigrcorrelation}
\end{figure}

Further insight can be gained when plotting (Fig.~\ref{f:rcorrelation})
the individual ratios 
$d_{\mathrm{NLTE}}/d_{\mathrm{Shkl}}$ as function of the PN radii
(given in Table~\ref{t:dist}). A strong correlation is present. A simple
(double logarithmic) linear fit results in 
\begin{equation}
\log\frac{d_{\mathrm{NLTE}}}{d_{\mathrm{Shkl}}} = 0.388 +0.935 \log 
				\frac{R}{\mathrm{pc}}\ .
\label{eq:fit}
\end{equation}
This indicates that the PN masses increase with increasing radii.
The trigonometric measurements essentially confirm this result
(Fig.~\ref{f:trigrcorrelation}). 
The effect of the 35\% shorter trigonometric distances in 
(Fig.~\ref{f:trigrcorrelation}) is small, because moderate changes of the 
distances essentially produce a shift parallel to our fit line. 

How does our result compare with other tests of the Shklovsky distance scale?
Ciardullo et al.\ (\cite{CBS99}) determined PN distances from main sequence
companions of the CSPN and concluded that statistical distance determinations
{\em overestimate} the PN distances. Stasi\'nska et al.\ (\cite{STA91}) and 
Pottasch \& Zijlstra (\cite{PZ92}) 
applied the Shklovsky method to PN in the bulge
of our Galaxy. Stasi\'nska et al.\ concluded that, besides considerable
scatter, the mean Shklovsky distance reproduces the distance of the galactic
bulge well, while Pottasch \& Zijlstra claimed that the Shklovsky distances
are systematically {\em too high}. Do these results contradict our findings
in Fig.~\ref{f:rcorrelation}?

Ciardullo et al.\ (\cite{CBS99}) derived spectroscopic 
parallaxes of main sequence
companions of 14 CSPN and compared their results (and 7 trigonometric
parallaxes) with four different statistical distance scales based on the
prescriptions of Cahn et al.\ (\cite{CKS92}), 
Maciel \& Pottasch (\cite{MP80}), van der
Steene \& Zijlstra (\cite{VSZ95}) and  Zhang (\cite{Zha95}).
These are variants of the Shklovsky method (Eq.~\ref{eq:Shkl}) in which the
ionized mass $M_{\mathrm{ion}}$ grows as a function of radius (Cahn et al.\
\cite{CKS92}; Maciel \& Pottasch \cite{MP80}; Zhang \cite{Zha95}) or of 
the radio brightness
temperature of the PN. Cahn et al.\ (\cite{CKS92})
assumed that the ionized mass of small, ionization bounded PNe grows with
radius until an upper limit is reached for larger density bounded PN.
This has the effect that these distance determinations for PNe in the sample
discussed by Ciardullo et al.\ (\cite{CBS99}) 
are essentially classical Shklovsky
distances with $M_{\mathrm{ion}} = 0.135M_\odot$. We performed a small
correction to the Cahn et al.\ (\cite{CKS92}) distances to transform them to 
the
common $M_{\mathrm{ion}} = 0.2M_\odot$ scale and added the Ciardullo et al.\
(\cite{CBS99}) data points to Fig.~\ref{f:trigrcorrelation}. 
Two conclusions can be drawn: 
\begin{enumerate}
\item the Ciardullo et al.\ results agree very well with the correlation
found for the spectroscopic distances in Fig.~\ref{f:rcorrelation},
\item the Ciardullo et al.\ distances are on average smaller than the Cahn
et al.\ distances, indeed. The reason is that the PNe investigated by
Ciardullo et al.\ are, on average, smaller than the PNe in our sample.
\end{enumerate}

PNe in the galactic bulge are very useful for several tests, because their
distance is known from the distance of the bulge. Stasi\'nska et al.\
(\cite{STA91})
and Pottasch \& Zijlstra (\cite{PZ92}) used this for tests of the Shklovsky
distance scale (both groups adopted a distance of 7.8\,kpc). Stasi\'nska et
al.\ (\cite{STA91}) found reasonable agreement of Shklovsky distance: their
distance distribution peaked between 8 and 9\,kpc with 77\% of all PNe in the 
range $8.5\pm 3.5$\,kpc. Pottasch \& Zijlstra (\cite{PZ92}) on the other hand
derived a peak value of 11.5\,kpc and a 75\% range from 7 to 17\,kpc. The main
difference between both works is that Pottasch \& Zijlstra used only fluxes and
angular diameters measured from radio observations, while Stasi\'nska et al.\
combined radio and optical flux data and preferred optical angular diameters. 

Since one wants to exclude foreground objects, PNe with an angular diameter
larger than $20''$ are excluded from bulge samples. This translates into a
radius $R = 0.41$\,pc. Thus the old and large PNe of our sample
(Table~\ref{t:dist}) are explicitly excluded from investigations of bulge PNe.
Another strong selection effect against old PNe might already be at work, 
because
of their low surface brightness combined with the large extinction. A lower
limit of the angular diameter of $\approx 1''$ is set by the need to resolve
the PNe. This constrains the radii of bulge PNe, which could be used to test
the Shklovsky method approximately to the range 
$0.04\,\mathrm{pc} < R < 0.41$\,pc. From Fig.~\ref{f:rcorrelation} we would
predict that the Shklovsky distances of bulge PNe are moderately too large,
in qualitative agreement with the Pottasch \& Zijlstra (\cite{PZ92}) result.
However, one should keep in mind that the stellar population of the galactic
bulge is quite different from the local population. Therefore one should be
aware that the properties of bulge PNe might be different from local samples.

The correlation given in equation~\ref{eq:fit} translates into a
mass-radius relation
\begin{equation}
M_{\mathrm{ion}} = 1.87 R^{2.3}
\label{eq:mass}
\end{equation}
with $M$ in $M_\odot$ and $R$ in pc. Equation~\ref{eq:mass} implies that
the ionized masses of old PN are quite high. For an old PN with
$R = 2$\,pc one derives $M_{\mathrm{ion}} = 9.3M_\odot$. 

It is likely that during the PN evolution the complete material of the
slow ($\approx$10\,km/s) AGB wind material is swept-up by the faster
($\approx$$20\ldots 40$\,km/s) expanding planetary
shell (Schmidt-Voigt \& K\"oppen \cite{SVK87}; 
Marten \& Sch\"onberner \cite{MS91})
and could finally be incorporated into the PN. However, $9M_\odot$ is much
larger than the mass a typical CSPN precursor ($M = 1{\ldots} 2M_\odot$)
can loose during it's evolution. There are some points which should be
considered when interpreting the mass-radius relation (Eq.~\ref{eq:mass}):
\begin{enumerate}
\item the mass-radius relation would be modified  if the filling factor
   $\epsilon$ (cf.\ Eq.~\ref{eq:Shkl}) changes during PN evolution,
\item many PN of our sample (Table~\ref{t:dist}) have a large angular
   diameter (several arcminutes are not uncommon) and the surface brightness 
   is very
   low. Thus it is often a challenging observational task to measure
   emission line fluxes of these objects. As a consequence the observational
   data provided in literature is often of low quality. We intend to determine
   reliable \hbeta\ fluxes of old PN from Table~\ref{t:dist} to improve
   this situation.
\item Old PN  show often signs of an interaction with the
   surrounding interstellar medium (Borkowski et al.\ \cite{BSS90}; Tweedy \&
   Kwitter \cite{TK96} and references therein). The PNe shape is distorted 
   and it is difficult to estimate the angular diameter $\Theta$.
\item On the other hand one may speculate that part of the ionized material
   is even swept-up interstellar matter. Note that an interstellar particle
   density of 1\,cm$^{-3}$ corresponds to $0.05M_\odot$/pc$^{-3}$. 
   If the interstellar
   matter corresponding to the nebular volume
   is incorporated into the nebula the mass of the $R=2$\,pc PN 
   is increased by $1.7M_\odot$. This might be an interesting topic for
   future investigations of the PNe interstellar matter interaction. 
   However, more high quality observational material is
   necessary before one can start to attack this question.
\end{enumerate}
If the mass-radius relation in Eq.~\ref{eq:mass} is confirmed one can use it
in principle to produce a new variant of a statistical distance scale.
However, the exponent ($\beta = 2.3$) is uncomfortable close to the value
$\beta = 2.5$ at which the method degenerates and unacceptable large errors
have to be expected (cf.\ e.g.\ Kwok \cite{Kwo00}, p.~177 f.).

\section{Conclusions}

We used the results of our model atmosphere analysis to compute
distances of the central stars and showed that 
our model atmosphere distance scale is in good agreement with measured
trigonometric parallaxes (after the effect introduced by biases has been taken
into account via a Monte Carlo simulation)
and distances derived from the interstellar
NaD lines. All these three independent methods of distance measurement
indicate that the widely used ``statistical'' distance scales of the
Shklovsky type are too short for old PNe. The most likely explanation
is an underestimate of the nebula masses. 

Estimates of the PNe space
density and birthrate, which are based on Shklovsky type distances,
therefore give too large values. If a more realistic distance scale is
applied, discrepancies between white dwarf and PNe birthrates are resolved.

\begin{figure}
\epsfxsize=8.5cm
\epsffile[18 40 750 580]{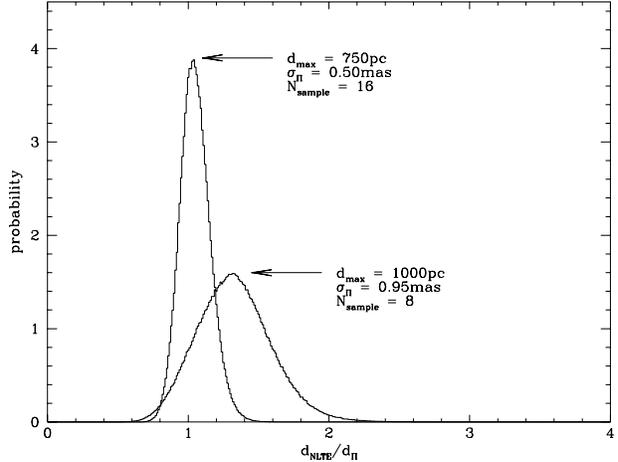}
\caption[]{Probability distribution for our standard model and for the 
possible future improvements discussed in the text}
\label{f:mcideal}
\end{figure}

Due to statistical uncertainties and biases of the trigonometric distances we
could test the spectroscopic distance scale
only on the 20\% level. Do we have to
wait for space missions like GAIA or SIM to improve this situation? A test
on the 10\% level might be achieved with state-of-the-art-techniques. Harris
et al.\ (\cite{HDM97}) announced that parallax measurement
accuracies of 0.5\,mas are within
reach, but this alone is not enough (cf.\ Table~\ref{t:mcvar}). However,
if a sample with $d_{\mathrm{NLTE}} < 750$\,pc is produced and the sample
size is doubled to $N_{\mathrm{sample}} = 16$ the bias $r_{\mathrm{MC}}$ is 
reduced to 4\% and the scatter to 10\% (the probability distribution is
compared with the present situation in Fig.~\ref{f:mcideal}). 
Thus, if we work on both fields, 
measuring of more accurate parallaxes and analyzing and selecting more 
candidates from spectroscopic investigations, this goal is within reach.

\acknowledgements
The author thanks Joachim K\"oppen for inspiring discussions and Detlef
Sch\"onberner, Uli Heber, and Klaus Werner for useful comments on 
previous drafts of this paper.

\appendix
\section{Monte Carlo simulation of the local PN population}
\label{s:mc}

Our Monte Carlo simulation of the local PN distribution proceeds in three
steps:
\begin{enumerate}
\item a star is created at a random position based on a simple 3D model of 
the galactic disk, 
\item a post-AGB age is randomly chosen and the absolute magnitude
	of the star is computed from a theoretical evolutionary track,
\item a scheme which describes the
selection of CSPNe for parallax measurements and the effects of measurement
errors is applied. 
\end{enumerate}
The parameters of our standard model are provided in
Table~\ref{t:mcstd} and the results of a parameter study which explores the 
effects of varying these parameters are listed
in Table~\ref{t:mcvar}.

\begin{table}
\caption[]{Standard parameters for the Monte Carlo simulation}
\label{t:mcstd}
\begin{tabular}{l@{= }r@{\,}l}
\multicolumn{3}{l}{Sample sizes:}\\
$N_{\mathrm{total}}$	&$10^5$&\\
$N_{\mathrm{sample}}$	&8&\\
\multicolumn{3}{l}{Sample selection:}\\
$V_{\mathrm{max}}$	&17.5&mag	\\
$d_{\mathrm{max}}$	&1000&pc\\	
$\Pi_{\mathrm{min}}$	&1.33&mas	\\
$\sigma_\Pi$		&0.95&mas\\	
$\sigma_{M_V}$		&0.75&mag\\
\multicolumn{3}{l}{Stellar evolution:}\\
$M_{\mathrm{CSPN}}$     &0.605&$M_\odot$\\
$T_{\mathrm{min}}$	&3000&yrs\\
$T_{\mathrm{max}}$	&50000&yrs\\
\multicolumn{3}{l}{Galactic model:}\\
$R_0$			&4500&pc\\
$H_{\mathrm{star}}$	&300&pc\\
$H_{\mathrm{dust}}$	&140&pc\\	
$\kappa_{\mathrm{dust}}$&1.0&mag/kpc	
\end{tabular}
\end{table}
\begin{table}
\caption[]{Influence of parameter variations on the results of the Monte
Carlo simulation}
\label{t:mcvar}
\begin{tabular}{lr@{\,}lr}
parameter		&\multicolumn{2}{c}{values}
&$d_{\mathrm{NLTE}}/d_\Pi$\\ \hline
std.\ model		&&		&1.32$\pm$0.25\\ \hline
$N_{\mathrm{sample}}$	&16&		&1.30$\pm$0.19\\
			&32&		&1.30$\pm$0.13\\
			&64&		&1.30$\pm$0.09\\
$V_{\mathrm{max}}$	&16.5&mag	&1.24$\pm$0.25\\
			&18.5&mag	&1.34$\pm$0.25\\ \hline
$d_{\mathrm{max}}$	&750&pc	&1.14$\pm$0.20\\
			&1500&pc	&1.50$\pm$0.38\\
			&3000&pc	&1.64$\pm$0.51\\ \hline
$\Pi_{\mathrm{min}}$	&1.5&mas	&1.37$\pm$0.26\\
			&2.0&mas	&1.51$\pm$0.30\\
			&3.0&mas	&1.51$\pm$0.35\\
			&4.0&mas	&1.32$\pm$0.30\\ 
			&5.0&mas	&1.22$\pm$0.24\\ 
			&7.0&mas	&1.16$\pm$0.21\\ 
			&10.0&mas	&1.14$\pm$0.20\\ \hline
$\sigma_\Pi$		&0.1&mas	&1.06$\pm$0.16\\
			&0.2&mas	&1.07$\pm$0.16\\ 
			&0.3&mas	&1.10$\pm$0.16\\ 
			&0.4&mas	&1.13$\pm$0.17\\ 
			&0.5&mas	&1.16$\pm$0.17\\ 
			&0.7&mas	&1.23$\pm$0.20\\ 
			&1.2&mas	&1.38$\pm$0.31\\ \hline
$\sigma_{M_V}$		&0.5&mag	&1.26$\pm$0.23\\
			&1.0&mag	&1.30$\pm$0.28\\ \hline
$T_{\mathrm{min}}$	&1000&yrs	&1.31$\pm$0.26\\
			&5000&yrs	&1.31$\pm$0.26\\ \hline
$T_{\mathrm{max}}$	&25000&yrs	&1.33$\pm$0.26\\
			&100000&yrs	&1.29$\pm$0.26\\ \hline
$H_{\mathrm{star}}$	&150&pc	&1.28$\pm$0.26\\
			&600&pc	&1.34$\pm$0.26\\ \hline
$H_{\mathrm{dust}}$	&60&pc		&1.32$\pm$0.25\\
			&210&pc	&1.31$\pm$0.25\\ \hline
$\kappa_{\mathrm{dust}}$&0.50&mag/kpc	&1.32$\pm$0.26\\
			&1.50&mag/kpc  &1.30$\pm$0.26\\
\end{tabular}
\end{table}

Our simple description of the galactic disk is based on the
Galaxy model of Bienaym\'e et al.\ (\cite{BRC87}). An exponential density law
with a scale height $H_{\mathrm{star}}$ of 300\,pc is adopted for the CSPN.
This corresponds to
a stellar population with an age of $3{\ldots} 5\cdot 10^9$\,yrs. We included
extinction by a dust component with  a scale height $H_{\mathrm{dust}}
=140$\,pc, which is the value appropriate for the interstellar matter. A
dust opacity at the position of the sun $\kappa_{\mathrm{dust}} = 1$\,mag/kpc
was chosen.
The
stellar and dust density decreases exponentially with the distance from the
galactic center and a scale length $R_0 = 4.5$\,kpc. For the distance of the
sun from the galactic center we adopted the standard value of 8.5\,kpc. 
Let us note that K\"oppen \& Vergely (\cite{KV98}) could successfully 
reproduce the extinction properties of galactic bulge PNe with our 
parameter values.

Our parameter study in Table~\ref{t:mcvar} shows that the influence of
particular values of the parameters of our ``galactic model'' are very small.
Due to the exponential decrease of stellar density with height above the
galactic plane the number of stars within a sphere with a given radius $R$
increases less then $R^3$. Since this lower the numbers of far away stars,
a lower bias is expected for lower values of the scale height. Extinction
introduces another selection against far away stars (through the limiting 
magnitude). However, as
Table~\ref{t:mcvar} proofs the effect of varying these parameters within
reasonable limits is quite small.

For a given post-AGB age the absolute magnitudes of CSPNe can be computed from
theoretical post-AGB tracks. In principle a mass distribution for the
CSPNe should be used. However, since the mass distributions of CSPN and white
dwarfs are quite narrow, we considered it sufficient to use only one
theoretical track for simplicity. Spectroscopic studies of CSPNe (Paper~IV),
white dwarfs (Bergeron et al.\ \cite{BSL92}, Napiwotzki et al.\ \cite{NGS99}) 
and an
investigation of PNe based on a distant independent method by Stasi\'nska
et al.\ (\cite{SGT97}) 
derived peak masses in the range $0.55{\ldots} 0.60 M_\odot$.
Thus we selected the $0.605M_\odot$ post-AGB track of Bl\"ocker (\cite{Blo95})
for our simulation. The post-AGB age was varied within a given interval
$T_\mathrm{min} {\ldots} T_\mathrm{max}$ which approximately reproduces the
$M_V$ distribution of our CSPNe. However the simulation results are quite 
insensitive to their particular values (cf.\ Table~\ref{t:mcvar}).

After ``producing'' a CSPN we had to simulate the selection for parallax
measurement and the distance determination with their measurement errors.
The following scheme was adopted:
\begin{enumerate}
\item we simulated a ``spectroscopic analysis'' with a random measurement
	error. We adopted a standard value for the scatter of the absolute
	magnitude determination of $\sigma_{M_V} = 0.75$ (Gaussian error
	distribution). This corresponds
	to the mean of the errors given in Table~\ref{t:dist}.
\item if the ``spectroscopic analysis'' indicates that the distance of the
	CSPN is below a given maximum value $d_{\mathrm{max}}$ the star
	was selected for ``parallax measurement''.
\item the parallax is ``measured'' with an Gaussian error distribution with 
        $\sigma_{\Pi}$.
	Our standard value of $\sigma_{\Pi}$ is 0.95\,mas. That is the mean
	error of the parallax measurements provided in Table~\ref{t:dist}.
	A parallax measurement is used for distance determination if the
	value is larger than a lower limit $\Pi_{\mathrm{min}}$. We chose
	$\Pi_{\mathrm{min}}=1.33$\,mas, corresponding to the value measured
	for \object{A\,74}.
\item this process was repeated until $N_{\mathrm{sample}}$
	distance determinations were performed. The standard value of
	$N_{\mathrm{sample}}$ is 8, the number of CSPNe with parallax
	measurements in	Table~\ref{t:dist}.
	The mean value of $d_\mathrm{NLTE}/d_\Pi$ was computed with the
	same method as used for our observed sample (cf.\ Section~2.2.).
\item to derive a proper statistic this simulation was repeated
	$N_\mathrm{total} = 10^5$ times. The best values given in
	Table~\ref{t:mcvar} and in Section~2.3 are the median values.
	$1 \sigma$ values were computed from the condition that they should
	include 68.3\% of the simulations. Results from repeated runs
	yielded results, which didn't deviate by more than 1\%.
\end{enumerate}
The effect of the most important input parameters, the 
threshold value $d_{\mathrm{max}}$ for the pre-selection of stars, the
parallax measurement error $\sigma_\Pi$ and the threshold parallax
$\Pi_{\mathrm{min}}$, is discussed in Section~2.2.


\begin{thebibliography}{}
\bibitem[1992]{AGG92}
Arenou F., Grenon M., G\'omez A. 1992, \aua\ 258, 104
\bibitem[1992]{BSL92}
Bergeron P., Saffer R.A., Liebert J. 1992, ApJ 394, 228
\bibitem[1987]{BRC87}
Bienaym\'e O., Robin A.C., Cr\'ez\'e M. 1987, \aua\ 180, 94
\bibitem[1952]{Bin52}
Binnendijk L. 1952, \apj\ 115, 428
\bibitem[1995]{Blo95}
Bl\"ocker T. 1995, \aua\ 299, 755
\bibitem[1990]{BSS90}
Borkowski K.J., Sarazin C.L., Soker N. 1990, \apj\ 360, 173
\bibitem[1992]{CKS92}
Cahn J.H., Kaler J.B., Stanghellini L. 1992, \auas\ 94, 399
\bibitem[1999]{CBS99}
Ciardullo R., Bond H.E., Sipior M.S., et al.\ 1999, \aj\ 118, 488
\bibitem[1982]{Dau82}
Daub C.T. 1982, \apj\ 260, 612
\bibitem[1995]{DSU95}
Dinerstein H.L., Sneden C., Uglum J. 1995, \apj\ 447, 262
\bibitem[1951]{Duk51}
Duke D. 1951, \apj\ 113, 100
\bibitem[1999]{GAL99}
Guti\'errez-Moreno A., Anguita C., Loyola P., Morena H. 1999, \pasp\ 111, 1163
\bibitem[1997]{HDM97}
Harris H.C., Dahn C.C., Monet D.G., Pier J.R 1997, in: IAU Symp.~180, 
	Planetary Nebulae, eds.\ H.J.~Habing \& H.J.G.L.M.~Lamers,
	Kluwer Academic Publ., Dordrecht, p.~40
\bibitem[1998]{HM98}
Hauck B., Mermilliod M. 1998, A\&AS 129, 431
\bibitem[1984]{HHJ84}
Heber U., Hunger K., Jonas G., Kudritzki R.P. 1984, A\&A 130, 119
\bibitem[1987]{IW87}
Ishida K., Weinberger R. 1987, \aua\ 178, 227
\bibitem[1995]{JVS95}
Jacoby G.H., Van de Steene G. 1995, \aj\ 110, 1285
\bibitem[1998]{KV98}
K\"oppen J., Vergely J.-L. 1998, \mnras\ 299, 567
\bibitem[2000]{Kwo00}
Kwok S. 2000, The origin and evolution of planetary nebulae, Cambridge
    University Press
\bibitem[1983]{Lut83}
Lutz T.E. 1983, in: Nearby stars and the Stellar Luminosity Function, IAU
    Coll.~76, eds.\ A.G.~Davis Philip, A.R.~Upgren, L.~Davis Press,
    Schenectady, p.~41
\bibitem[1973]{LK73}
Lutz T.E., Kelker D.H. 1973, \pasp\ 85, 573
\bibitem[1980]{MP80}
Maciel W.J., Pottasch S.R. 1980, \aua\ 88, 1
\bibitem[1991]{MS91}
Marten H., Sch\"onberner D. 1991, \aua\ 248, 590
\bibitem[1988]{MKH88}
M\'endez R.H., Kudritzki R.P., Herrero A., Husfeld D., Groth H.G. 1988,
	A\&A 190, 113
\bibitem[1999]{Nap99}
Napiwotzki R. 1999, \aua\ 350, 101 (Paper~IV), astro-ph/9908181
\bibitem[1995]{NS95}
Napiwotzki R., Sch\"onberner D. 1995, \aua\ 301, 545 (Paper~III)
\bibitem[1999]{NGS99}
Napiwotzki R., Green P.J., Saffer R.A. 1999, \apj\ 517, 399, astro-ph/9901027
\bibitem[1993]{NSW93}
Napiwotzki R., Sch\"onberner D., Wenske V. 1993, \aua\ 268, 653
\bibitem[1993]{PHD93}
Pier J.R., Harris H.C., Dahn C.C., Monet D.G. 1993, in: IAU Symp.~155, 
	Planetary nebulae, eds.\ R.~Weinberger \& A.~Acker, Kluwer Academic 
	Publ., Dordrecht, p.~175
\bibitem[1984]{Pot84}
Pottasch S.R. 1984, Planetary Nebulae, Reidel Publ., Dordrecht
\bibitem[1996]{Pot96}
Pottasch S.R. 1996, \aua\ 307, 561
\bibitem[1998]{PA98}
Pottasch S.R., Acker A. 1998, \aua\ 329, L5
\bibitem[1992]{PZ92}
Pottasch S.R., Zijlstra A.A. 1992, \aua\ 256, 251
\bibitem[1992]{PTV92}
Press W.H., Teukolsky S.A., Vetterling W.T., Flannery B.P. 1992,
	Numerical Recipes, Cambridge University Press
\bibitem[1950]{Ram50}
Ramsey J. 1950, \apj\ 111, 434
\bibitem[1995]{Sau95}
Saurer W. 1995, \aua\ 297, 261
\bibitem[1992]{SSM92}
Schaller G., Schaerer D., Meynet G., Maeder A. 1992, A\&AS 96, 269
\bibitem[1987]{SVK87}
Schmidt-Voigt M., K\"oppen J. 1987, \aua\ 174, 223
\bibitem[1956]{Shk56}
Shklovsky I.S. 1956, Sov.\ Astron.\ J.\ 33, 315
\bibitem[1987]{Smi87}
Smith H.Jr. 1987, \aua\ 188, 233
\bibitem[1991]{STA91}
Stasi\'nska G., Tylenda R., Acker A., Stenholm B. 1991, \aua\ 247, 173
\bibitem[1997]{SGT97}
Stasi\'nska G., G\'orny S.K., Tylenda R. 1997, A\&A 327, 736
\bibitem[1953]{TW53}
Trumpler R.J., Weaver H.F. 1953, Statistical Astronomy, University of
	California Press, Berkeley
\bibitem[1996]{TK96}
Tweedy R.W., Kwitter K.B. 1996, \apjs\ 107, 255
\bibitem[1995]{VSZ95}
Van de Steene G.C., Zijlstra A.A. 1995, \aua\ 293, 541
\bibitem[1991]{Wei91}
Weidemann V. 1991, in: White dwarfs, eds.\ G.~Vauclair \& E.~Sion,
	Kluwer, Dordrecht, p.~67 
\bibitem[1995]{Zha95}
Zhang C.Y. 1995, \apjs\ 98, 659
\end{thebibliography}
\end{document}